\documentclass[11pt,twocolumn]{article}
\usepackage[utf8]{inputenc}
\usepackage{fullpage}
\usepackage{aas_macros}
\usepackage{hyperref}
\usepackage{dblfloatfix}
\usepackage{mathtools}

\usepackage[font=footnotesize,labelfont=bf]{caption}
\usepackage{wrapfig}
\def\micron{$\mu$m}
\def\etal{{\em et al.}}

\usepackage{fancyhdr}
\usepackage[compact]{titlesec}
\titlespacing{\section}{0pt}{*1}{*2}
\titlespacing{\subsection}{0pt}{*1}{*2}
\titlespacing{\subsubsection}{0pt}{*.5}{*2}
\titlespacing{\paragraph}{0pt}{*.5}{*2}
\usepackage{mdwlist}

\usepackage[decimalsymbol=.,separate-uncertainty = true,multi-part-units=single,range-phrase=-]{siunitx}

\DeclareSIUnit\inch{''}

\def\etal{{\em et al.}}

\usepackage{ifpdf}

\ifpdf
\usepackage[pdftex]{graphicx}
\else
\usepackage{graphicx}
\fi

\ifpdf
\DeclareGraphicsExtensions{.pdf, .jpg, .tif}
\else
\DeclareGraphicsExtensions{.eps, .jpg}
\fi

\begin{document}

\onecolumn
\pagestyle{empty}
\raggedright
\huge
Astro2020 APC White Paper \linebreak

Optical and Near-IR Microwave Kinetic Inductance Detectors (MKIDs) in the 2020s\linebreak
\normalsize

\textbf{Principal Author:}

Name: Benjamin A. Mazin	
 \linebreak						
Institution: University of California Santa Barbara  
 \linebreak
Email: bmazin@ucsb.edu
 \linebreak
Phone: (805)893-3344
 \linebreak
 
\textbf{Co-authors:}
Jeb Bailey, Jo Bartlett, Clint Bockstiegel, Bruce Bumble, Gregoire Coiffard, Thayne Currie, Miguel Daal, Kristina Davis, Rupert Dodkins, Neelay Fruitwala, Nemanja Jovanovic, Isabel Lipartito, Julien Lozi, Jared Males, Dimitri Mawet, Seth Meeker, Kieran O’Brien, Michael Rich, Jenny Smith, Sarah Steiger, Noah Swimmer, Alex Walter, Nick Zobrist, and Jonas Zmuidzinas

\pagebreak
\pagestyle{fancy}
\setcounter{page}{1}
\twocolumn


\section{Introduction}  

Optical and near-IR (OIR) Microwave Kinetic Inductance Detectors, or MKIDs, are superconducting photon counting detectors capable of measuring the energy and arrival time of individual OIR photons without read noise or dark current. They are widely considered to be one of the most promising emerging OIR astronomical detectors.  In this whitepaper we will discuss the current status of MKIDs and MKID-based instruments, and lay out the exciting future they will enable with sufficient support from the community.

\section{Key Science Goals and Objectives}

MKIDs in the OIR are usually thought of as noise-free integral field spectrographs (IFSs) on a chip (\S\ref{sec:lowspec}). However, this is only the simplest use of an MKID.  Using them in more complex instruments opens up new possibilities such as large survey instruments (\S\ref{sec:supermos}) and multiplexed high resolution spectrometers (\S\ref{sec:highspec}). 

\subsection{Low Resolution MKID Integral Field Spectrographs}
\label{sec:lowspec}

MKID focal planes have so far been incorporated into two main classes of OIR instruments.  First, MKIDs have been used in seeing-limited IFSs for observations of time variable and low surface brightness objects with the ARCONS instrument~\cite{2013PASP..125.1348M}.  This has led to a number of interesting scientific results~\cite{2013ApJ...779L..12S,2014MNRAS.tmp..307S,Strader:2016he,2017ApJ...850...65C}, but a true facility class instrument is needed to make further progress.  One such proposal is for KRAKENS, an instrument of unprecedented capabilities for the Keck I bent Cassegrain port. KRAKENS will be a MKID-based, low resolution, high sensitivity IFS with at least 30,600 (170${\times}$180) pixels, a 42.5''${\times}$45'' field of view (FOV), spectral resolution $R=\lambda/\Delta \lambda{\approx}$40 at 0.4~$\mu$m, and the wide 0.38--1.35~$\mu$m bandwidth inherent in MKIDs.  This FOV will place stable calibration stars on the array in nearly every field for accurate photometry, guiding, and software tip/tilt corrections, as well as enabling observations of larger objects like nearby galaxies and galaxy clusters.  

The second class of IFS instruments using MKIDs have been for high contrast direct imaging of exoplanets.  For this application the near-IR sensitivity, lack of read noise, and fast readout enable active speckle suppression~\cite{2014PASP..126..565M} as well as extremely effective time-based speckle suppression using stochastic speckle discrimination~\cite{2019arXiv190603354W}.  Existing MKID instruments for this application include DARKNESS for the Palomar 200"~\cite{2018PASP..130f5001M}, MEC for Subaru SCExAO, and the PICTURE-C NASA balloon.

In the next decade we hope to use MKID backends for the Planetary Systems Imager (PSI) for TMT.  PSI is a modular instrument suite centered on a high-performance adaptive optics (AO) system that enables a broad range of sensing and characterization capabilities, with a particular focus on high contrast applications.  Built on a core capability of wavefront control and starlight suppression, PSI combines science backends providing imaging, polarimetry, integral field spectroscopy, and high-resolution spectroscopic capabilities across a wide range of wavelengths (0.6--5\,\micron, as well as a thermal channel at 10\,\micron). This instrument is very well suited to address major questions in the formation and evolution of planetary systems that will be relevant in the ELT era.  

\subsection{Superconducting Multiobject Spectrographs (SuperMOS)}
\label{sec:supermos}

In a conventional multi-object spectrograph, like DEIMOS on Keck~\cite{2003SPIE.4841.1657F}, a mask is inserted at the focal plane so that light from the targets may pass through the slits (or apertures), but background sky and other nearby source photons are blocked to reduce sky noise and contamination.  After passing through the mask, a dispersive element such as a diffraction grating or prism is used to spread the light as a function of wavelength on a detector.  In a SuperMOS, we use the same mask-based approach to reduce sky background and contamination from other sources, but require no dispersive element, instead using the intrinsic spectral resolution of the MKID detectors.  Since each MKID pixel provides moderate spectral resolution the focal plane is used much more efficiently, yielding a simple and compact system.  

An example of of this type of instrument is the Giga-$z$~\cite{2013ApJS..208....8M} concept.  It is enabled both by the inherent energy resolution and especially by the large pixel counts possible with the MKIDs.  It is currently envisioned as an instrument for the Cassegrain or Naysmith focus of a dedicated 4-m class telescope.  In order to cover 20,000 square degrees in a reasonable amount of time, we use a one square degree focal plane.  This square degree field of view is divided among the 100,000 detectors, each fed by a macropixel covering 10''$\times$10'' of the sky.  Galaxy number counts in $I$ band to 24.5$^{th}$ magnitude~\cite{2015MNRAS.451.4238S} ensure that most macropixels (80\,-\,100\%) will contain a galaxy at each pointing.  A mask cut using pre-existing LSST (or earlier Dark Energy Survey) imaging allows the light from one celestial source per macropixel to fall onto a square microlens array (99\% fill factor), focusing the light onto the corresponding large plate scale MKID located directly below.  Depending on the specific camera design, a reimaging system incorporating demagnification would likely be required between the mask and the microlens array.  

A two year survey with the Giga-$z$ instrument would be capable of returning 30--60 spectral channels on nearly 2 billion galaxies down to m$_I{\sim}24.5$.  The wider wavelength coverage dramatically reduced the number of catastrophic failures in photo-z determination.  Giga-$z$ vastly improving the science return of LSST. 

\subsection{High Resolution Multiobject MKID Spectrographs (HRMOS) }
\label{sec:highspec}

High resolution spectrographs with spectral resolution R$>$5,000 are one of the most important tools for nearly every discipline in astronomy, but the design of high resolution spectrographs has been nearly static for the last several decades.  A high resolution spectrograph usually contains a collimator, echelle grating, cross disperser, and finally a camera and science detector.  The 2-d nature of the resulting image (echellogram) makes long slits practical, but along with detector read noise, has prevented the high number of simultaneous observations of different objects (high muiltiplex) seen in lower (R$<$5,000) resolution spectrographs using image slicers, fibers, and lenslets.

A radical new type of high resolution multiobject spectrograph (HRMOS) is enabled by using MKID focal planes.  MKIDs can determine the energy of each arriving photon without read noise or dark current, and with high temporal resolution.  If MKIDs are used it allows a conventional echelle spectrograph to be constructed without the cross disperser.  The energy resolution of the MKID allow us to determine which echelle order the photon came from~\cite{2003MNRAS.344...33C,2014SPIE.9147E..0GO}.  This drastically simplifies the instrument and allows the simultaneously measurement of the spectra of the output of many fibers using a series of parallel linear strips of MKIDs.  The lack of read noise from the MKIDs mean there is no penalty for measuring at a high spectral resolution and then re-binning to lower resolution for better photon statistics. This design could allow for future facility-class instrument working in the visible with ${>}100$ fibers inputs and R${>}$50,000. 

\begin{figure}
\begin{center}
\vspace{-0.0in}
\includegraphics[width=1.0\columnwidth]{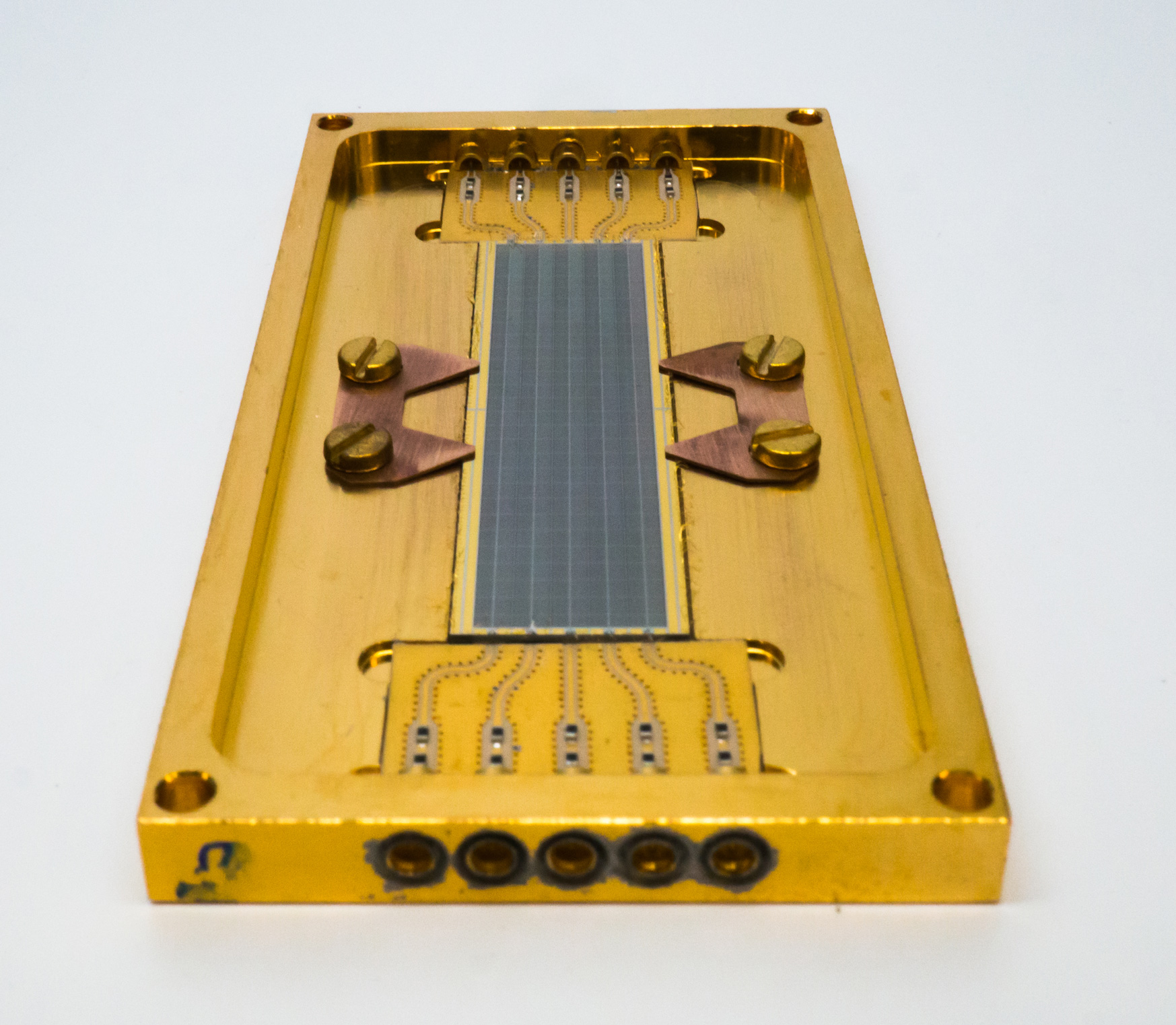}
\end{center}
\vspace{-.2in}
\caption{A 2048 $\times$ 5 pixel MKID array with the strips separated by 2 mm mounted in a microwave box.}
\label{fig:S1}
\vspace{-.1in}
\end{figure}

This development is enabled by a newly developed linear MKID array that has 2048$\times$N MKIDs, which contain 2048 MKIDs in each of N long linear strips, shown in Figure~\ref{fig:S1}. This arrangement allows high quantum efficiency without the use of microlenses. 

\subsubsection{HRMOS Science: High Dispersion Coronagraphy}
\label{sec:HDC}

\begin{figure}
\begin{center}
\vspace{-0.0in}
\includegraphics[width=0.95\columnwidth]{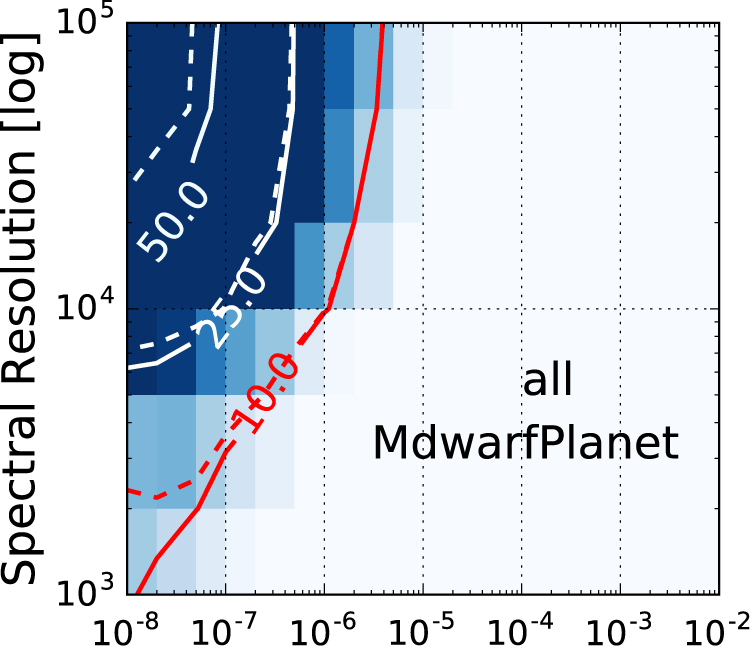}
\end{center}
\vspace{-.2in}
\caption{From Wang \etal~\cite{2017AJ....153..183W}: Results of a J-band simulation of an Earth-like planet around a M-dwarf star at 5 pc, showing cross-correlation S/N contours for given expected 30-m class telescope instrument spectral resolutions and starlight suppression levels. “All” means that all molecular lines are used. This plot shows that the HDC technique on a 30-m telescope can feasibly yield a S/N$>$25 detection of molecular lines.}
\label{fig:HDC}
\vspace{-.1in}
\end{figure}

A MKID HRMOS would be especially useful in the search for molecules in the spectra of planets around nearby stars using the cross-correlation technique~\cite{2010Natur.465.1049S,2013Sci...339.1398K,Barman:2015dy} behind a coronagraph~\cite{2015A&A...576A..59S}.  Known as High Dispersion Coronagraphy (HDC), individual spectra are cross-correlated with known molecular spectra (which cannot be present in the stray light from the star), allowing a significant increase (up to an extra factor of $10^{-4}$) raw contrast.  The combination of the extreme adaptive optics, advanced coronagraphs with small inner working angles, and the cross-correlation technique could potentially provide the extreme contrast ($<10^{-8}$ after post processing) needed for the detection of Earth analogues with 30-m telescopes~\cite{2017AJ....153..183W}, as shown in Figure~\ref{fig:HDC}.

In this HRMOS application, 100 fibers would be arranged into an IFU to sample the area around the star from roughly 1.5--7 $\lambda/D$ at the focal plane of a high contrast extreme AO instrument like the Planetary Systems Instrument (PSI) that was recently selected as the highest priority second generation instrument concept for TMT.   Having 100 fibers instead of just one or two in a conventional spectrograph vastly increases the utility of the HDC technique, as it goes from being only a follow-up technique (you have to find the exoplanet before you can put a fiber on it!) to being a simultaneous discovery and characterization technique.  

Since habitable zone planets will be extremely faint (Proxima Cen b is likely around 25$^{th}$ magnitude), the signal to noise ratio in the high resolution spectra is extremely low, making read noise and dark current-free detectors absolutely vital.  Even if you could get enough semiconductor pixels assembled to take the 100 simultaneous spectra around the 1~\micron~sweet spot~\cite{2017AJ....153..183W} required to find these planets, the read noise and dark current of even the most promising near-IR semiconductor detectors (like SAPHIRA, currently only available in a non-buttable 320${\times}$256 pixel format) will likely overwhelm the signal~\cite{Lovis:2017kt}. \emph{A HRMOS on TMT would have an excellent chance of both discovering the nearest planets and detecting biosignature gases in their atmospheres}. 

\subsubsection{HRMOS Science: Stellar Populations}

A MKID spectrograph without a cross disperser would provide two very significant gains for stellar spectroscopy ---  boosting efficiency by ${\sim}$40\% by removing the cross disperser and providing zero noise photon counting.   One can roughly estimate that the two advantages would enable one to reach S/N~30 at g=21 in 10h with TMT ---  a red giant at the distance of the Andromeda galaxy, (m-M)=24.5.  This brings genuine high resolution spectroscopy to the Andromeda galaxy and its satellites.

A MKID HRMOS would fill a critical gap in the TMT instrumentation suite due to the lack of an optical high resolution spectrograph.  Indeed, it would offer multi-object capability which would be unique even over the limited NFIRAOS (TMT's first light AO system) 2' FOV, or potentially even larger with seeing limited fiber injection mechanisms.  This opens up to TMT the subject area of chemical evolution in external galaxies.   The potential for a high resolution photon-counting spectrograph to do great work can be seen e.g. in McWilliam \etal~\cite{1995AJ....109.2757M} --- a paper on spectroscopy of metal poor stars using S/N$\sim$30 spectra with 709 citations. This paper could be repeated, but now for field stars in the halos of Local Group galaxies such as M31, M33, NGC 205, NGC 147, etc. As it becomes clear that a significant fraction of our Milky Way halo results from some kind of collision history~\cite{Helmi:2018hk} the study of halo populations in other galaxies becomes vital.

\begin{figure}
\begin{center}
\vspace{-0.15in}
\includegraphics[width=0.95\columnwidth]{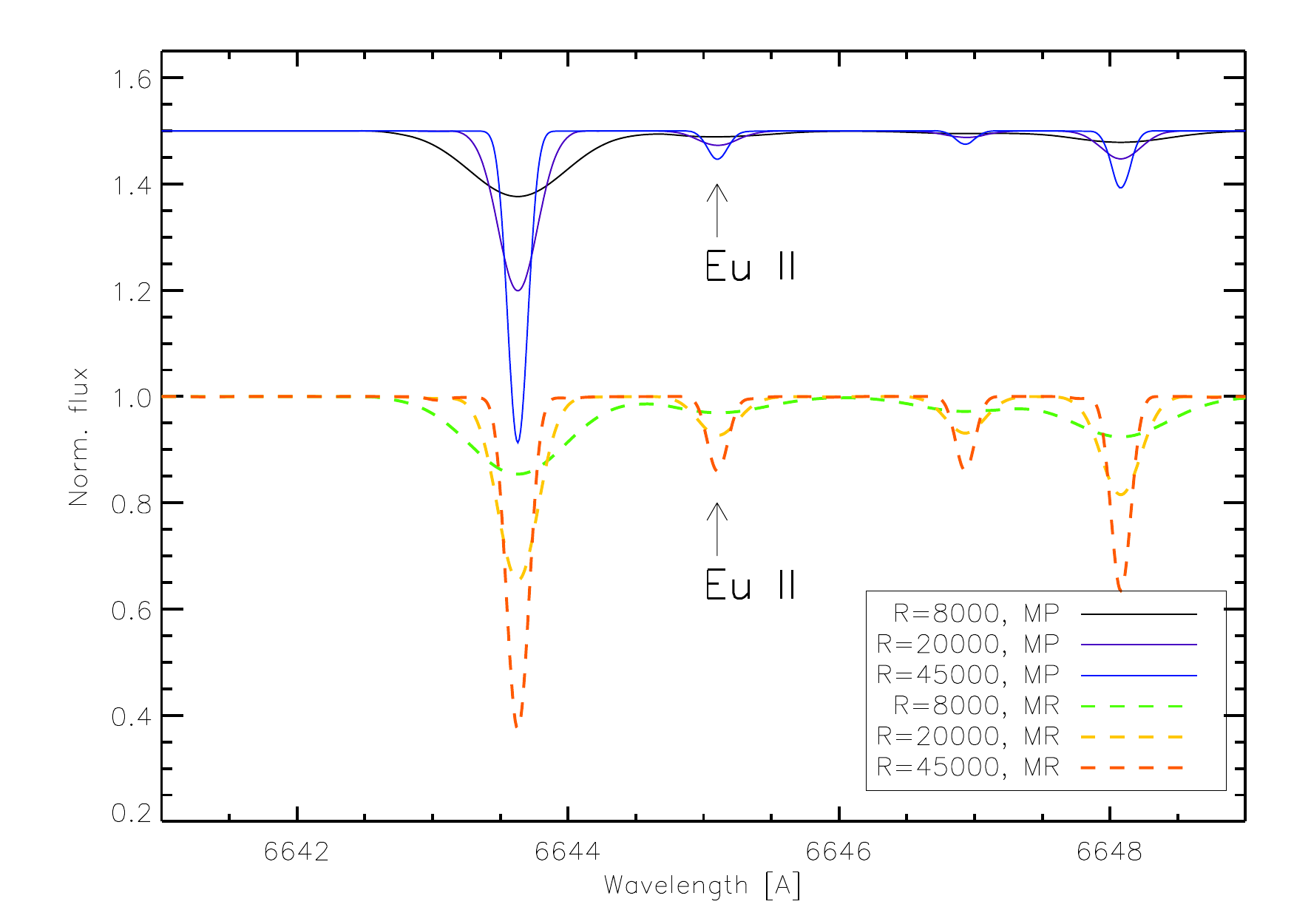}
\end{center}
\vspace{-.25in}
\caption{The need for higher resolution to measure the critical
r-process element Eu.   Eu is one of the few elements that
can only be produced via the r-process.  The Eu II 6645 A line is shown
at R=8000, 20000, and 45000 for a star with [Fe/H]=-2 and
a star with Solar metallicity.  The Eu II line is not measurable
at R=8,000, the currently planned resolution for TMT WFOS.}
\label{fig:stars}
\vspace{-.15in}
\end{figure}

At present, the Andromeda dwarf spheroidal galaxies, globular clusters, and halo field stars are well explored in terms of kinematics, but abundance determination at R$\sim8000$ is less reliable.  Medium resolution is problematic for abundance determination, as most measurable lines of interest would be on the flat part of the curve of growth. While some progress has been made for constraints on [Fe/H], there remain no reliable measurements of elemental composition for individual Andromeda halo or dwarf spheroidal stars.  
Some important lines, like the r-process element Eu, cannot be measured at R=8000. Figure~\ref{fig:stars}, courtesy of E. Kirby, shows this for a metal poor and metal rich star.  The widely used Eu II 6645 line requires high spectral resolution to measure.  

An MKID HRMOS on TMT would transform study of the chemical evolution of Local Group galaxies.  Are the most metal poor giants in M31 similar to those in the Milky Way?  Are there populations of carbon-enhanced metal poor stars?   Are abundance trends in the M31 globular cluster system similar to those found in its halo?  If the M31 halo formed via  different process, can we see that history in its chemical evolution?
Do M31 globular clusters contain multiple stellar populations, perhaps associated with different helium abundance, as is widely observed to be the case for clusters in the Milky Way?  There are also classes of M31 small stellar systems, such as the extended globular clusters, which have no Milky Way counterparts.  

Even among the Milky Way satellites the ultra-faint dwarf spheroidals have been difficult to study because they are so poorly populated that only subgiant and main sequence stars are present.  This places many of their members fainter than g=20, rendering them expensive to access with high resolution spectroscopy by present-day 8--10 m telescopes. 

As the controversy over the nature and structure of dark matter halos grows, the push to work on ever more distant systems will as well.  Most of these systems are $<2$ arcmin in size, so a HRMOS behind NFIRAOS could deliver hundreds of radial velocities for stars and globular clusters in M81 satellites and some ultra-diffuse galaxies. The goal will be to collect internal dynamics for a wide range of small stellar systems.

The integrated light of individual globular clusters in external galaxies could be used to derive abundances for individual elements out to the Virgo cluster and beyond with good accuracy.  At present, Fe , Ca, Si, and Ba derived from integrated light spectra give good agreement with individual stellar measurements~\cite{McWilliam:2008ux,2017ApJ...834..105C} but this technique requires R $\sim30,000$ spectra over a wide wavelength range.  These techniques may shed light on chemical evolution and populations of the halos for galaxies spanning a wide range in luminosity and Hubble type.

\section{Microwave Kinetic Inductance Detectors}

\label{sec:MKID}

Low temperature detectors (LTDs), with operating temperatures on the order of 100~mK, are currently the preferred technology for astronomical observations over most of the electromagnetic spectrum, notably in the far infrared through millimeter (0.1--3~mm)~\cite{Bintley:2010gz,Niemack:2008gk,Carlstrom:2011ik,2018A&A...609A.115A}, X-ray~\cite{Kelley:2009ce,Ulbricht:2015kz}, and gamma-ray~\cite{Doriese:2007hya} wavelength ranges.  In the important UVOIR (0.1--5~$\mu$m) wavelength range a variety of detector technologies based on semiconductors, backed by large investment from both consumer and military customers, has resulted in detectors for astronomy with large formats, high quantum efficiency, and low readout noise.  These detectors, however, are fundamentally limited by the large band gap of the semiconductor which restricts the maximum detectable wavelength (1.1 eV for silicon), and by thermal noise sources from their relatively high ($\sim$100~K) operating temperatures.  LTDs allow the use of superconductors with gap parameters over 1000 times lower than semiconductors.  This difference enables a leap in capabilities.  A superconducting detector can count single photons with no false counts while determining the energy (to several percent or better) and arrival time (to a microsecond) of the photon --- the optical analog of an X-ray calorimeter.  It can also have much broader wavelength coverage since the photon energy is always much greater than the gap energy.  While a CCD is limited to about 0.35--1~$\mu$m, the LTDs described here are in principle sensitive from 0.1~$\mu$m in the UV to greater than 5~$\mu$m.

\begin{figure}
\begin{center}
\vspace{-0.1in}
\includegraphics[width=1.0\columnwidth]{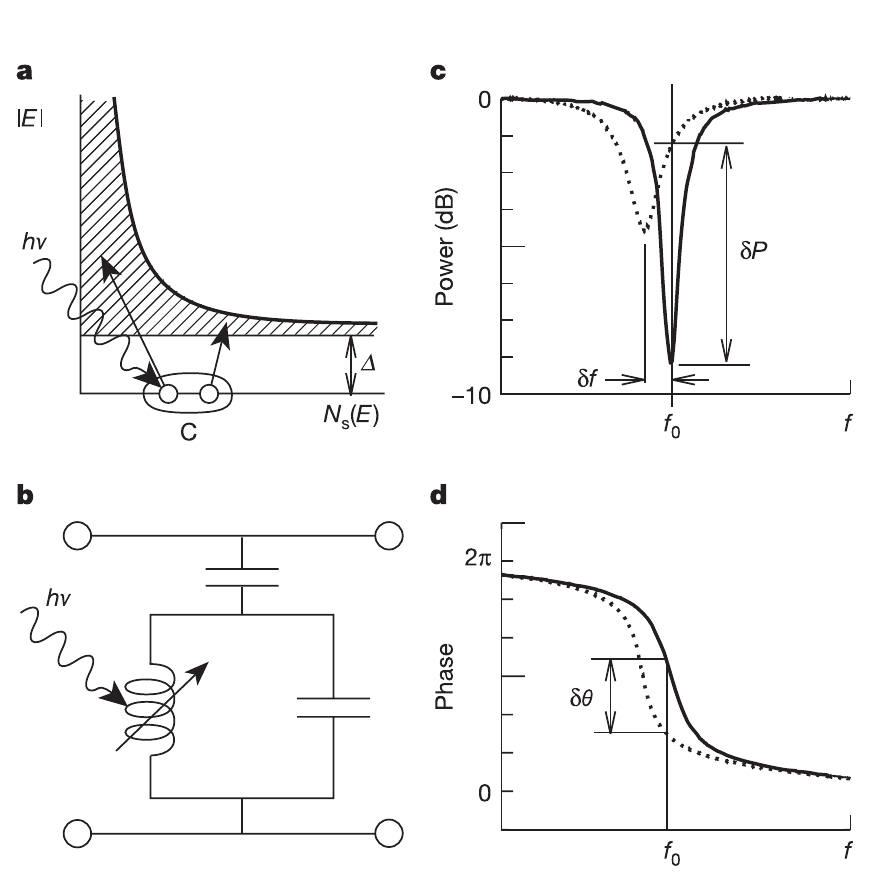}
\end{center}
\vspace{-.2in}
\caption{The basic operation of an MKID, from \cite{2003Natur.425..817D}. (a) Photons with energy $h\nu$ are absorbed in a superconducting film, producing a number of excitations, called quasiparticles.  (b) To sensitively measure these quasiparticles, the film is placed in a high frequency planar resonant circuit.  The amplitude (c) and phase (d) of a microwave excitation signal sent through the resonator.  The change in the surface impedance of the film following a photon absorption event pushes the resonance to lower frequency and changes its amplitude.  If the detector (resonator) is excited with a constant on-resonance microwave signal, the energy of the absorbed photon can be determined by measuring the phase and amplitude shifts.} 
\label{fig:detcartoon}
\vspace{-.1in}
\end{figure}

Superconducting UVOIR detectors have been pursued in the past with two technologies, Superconducting Tunnel Junctions (STJs)~\cite{Martin:2006dx,Hijmering:2008kn} and Transition Edge Sensors (TESs)~\cite{Romani:2001fw,Burney:2006ds}.  While both of these technologies produced functional detectors, they are limited to single pixels or small arrays due to the lack of a credible strategy for wiring and multiplexing large numbers of detectors, although recently there have been proposals for larger TES multiplexers~\cite{2010ApPhL..96p3509N}.

MKIDs \cite{2003Natur.425..817D,Mazin:2012kl} are a newer cryogenic detector technology that can be easily multiplexed into large arrays.  The ``microwave'' in MKIDs comes from their use of frequency domain multiplexing~\cite{mattthesis} at microwave frequencies (0.1--20 GHz) which allows thousands of pixels to be read out over a single microwave cable. The UVOIR lumped element~\cite{Doyle:2008gc} MKID arrays that we have developed have significant advantages over semiconductor detectors.  They can count individual photons with no false counts, determine the energy and arrival time of every photon with good quantum efficiency, and their physical pixel size is well matched to large telescopes.  These capabilities enable powerful new astrophysical instruments usable from the ground and space. 

\begin{figure*}
\begin{center}
\vspace{0in}
\includegraphics[width=2.0\columnwidth]{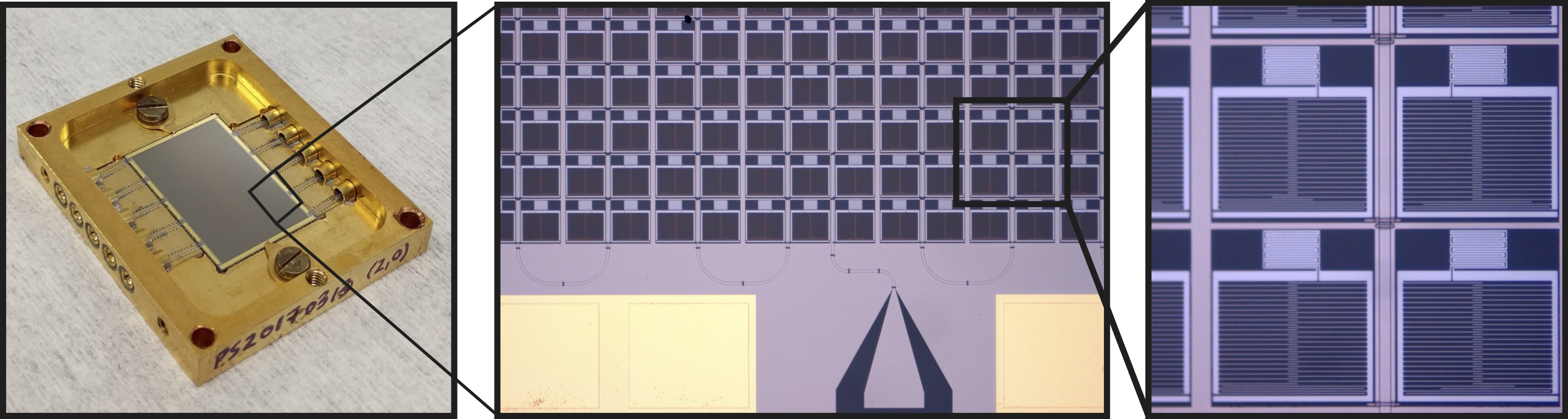}
\end{center}
\vspace{-.2in}
\caption{Left: A 10,000 MKID PtSi on Sapphire array for DARKNESS mounted in its microwave package. Center: Detail image of the array showing the CPW transmission line with bond pad for one feedline. Right: Further detail of several MKID pixels. The densely meandered patches at the top of each pixel are the photosensitive inductors, and the large sparse sections are the interdigitated capacitors used to tune each MKID to a unique resonant frequency.  A microlens arrays is used to boost the effective fill factor to 90\%.} 
\label{fig:SCI4}
\vspace{-.1in}
\end{figure*}

MKIDs work on the principle that incident photons change the surface impedance of a superconductor through the kinetic inductance effect~\cite{Mattis:1958vo}.  The kinetic inductance effect occurs because energy can be stored in the supercurrent (the flow of Cooper Pairs) of a superconductor.  Reversing the direction of the supercurrent requires extracting the kinetic energy stored in it, which yields an extra inductance term in addition to the familiar geometric inductance.  The magnitude of the change in surface impedance depends on the number of Cooper Pairs broken by incident photons, and hence is proportional to the amount of energy deposited in the superconductor. This change can be accurately measured by placing a superconducting inductor in a lithographed resonator, as shown in Figure~\ref{fig:detcartoon}.  A microwave probe signal is tuned to the resonant frequency of the resonator, and any photons which are absorbed in the inductor will imprint their signature as changes in phase and amplitude of this probe signal.  Since the quality factor $Q$ of the resonators is high and their microwave transmission off resonance is nearly perfect, multiplexing can be accomplished by tuning each pixel to a different resonant frequency with lithography during device fabrication.  A comb of probe signals can be sent into the device, and room temperature electronics can recover the changes in amplitude and phase without significant crosstalk~\cite{2012RScI...83d4702M}.

\begin{figure}
\begin{center}
\vspace{0in}
\includegraphics[width=1.0\columnwidth]{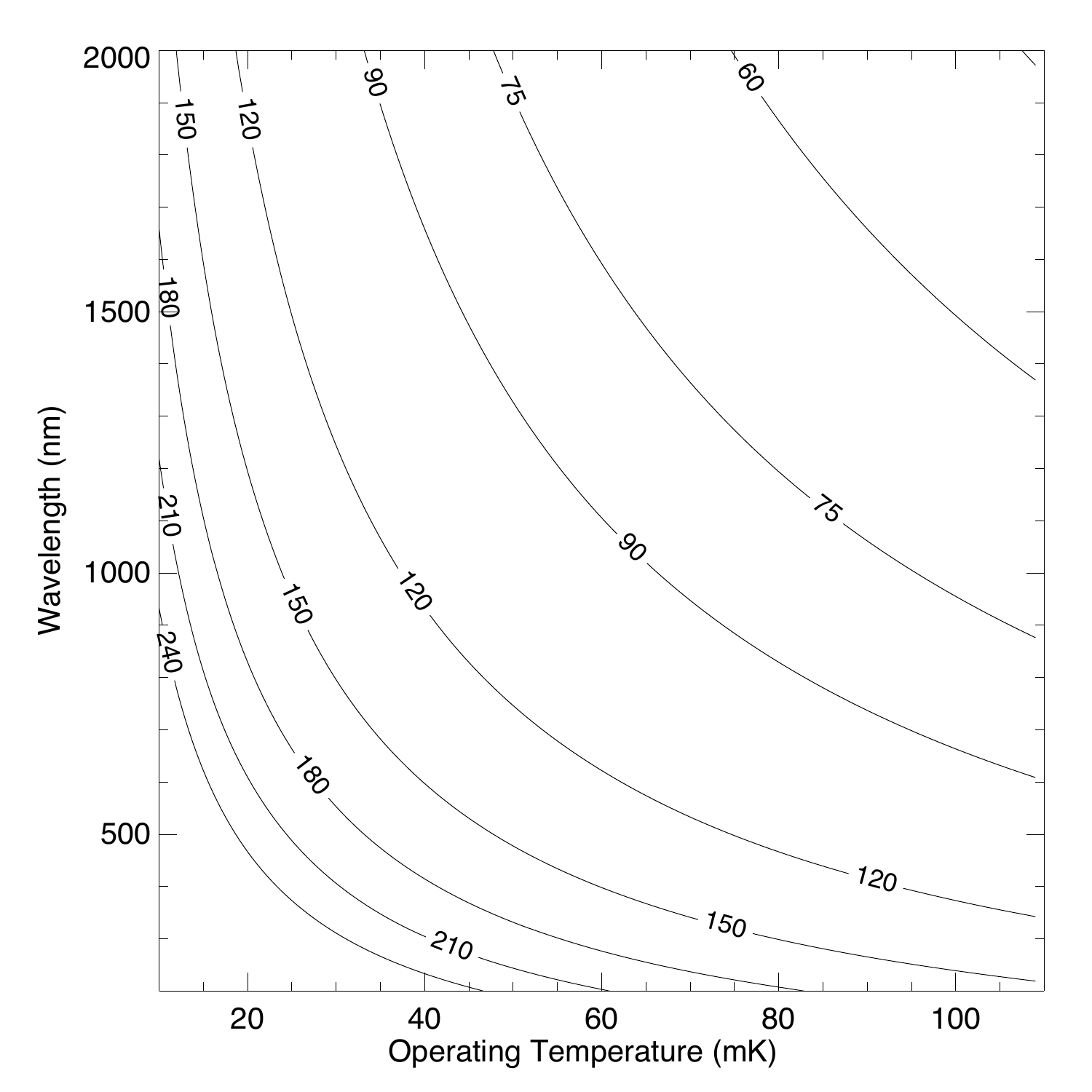}
\end{center}
\vspace{-.2in}
\caption{A plot of the Fano Limit as a function of operating temperature (assuming the MKID material has $T_c = 8 T_{op}$) and photon wavelength.} 
\label{fig:fano}
\vspace{-.1in}
\end{figure}

\begin{figure}
\begin{center}
\vspace{0in}
\includegraphics[width=0.9\columnwidth]{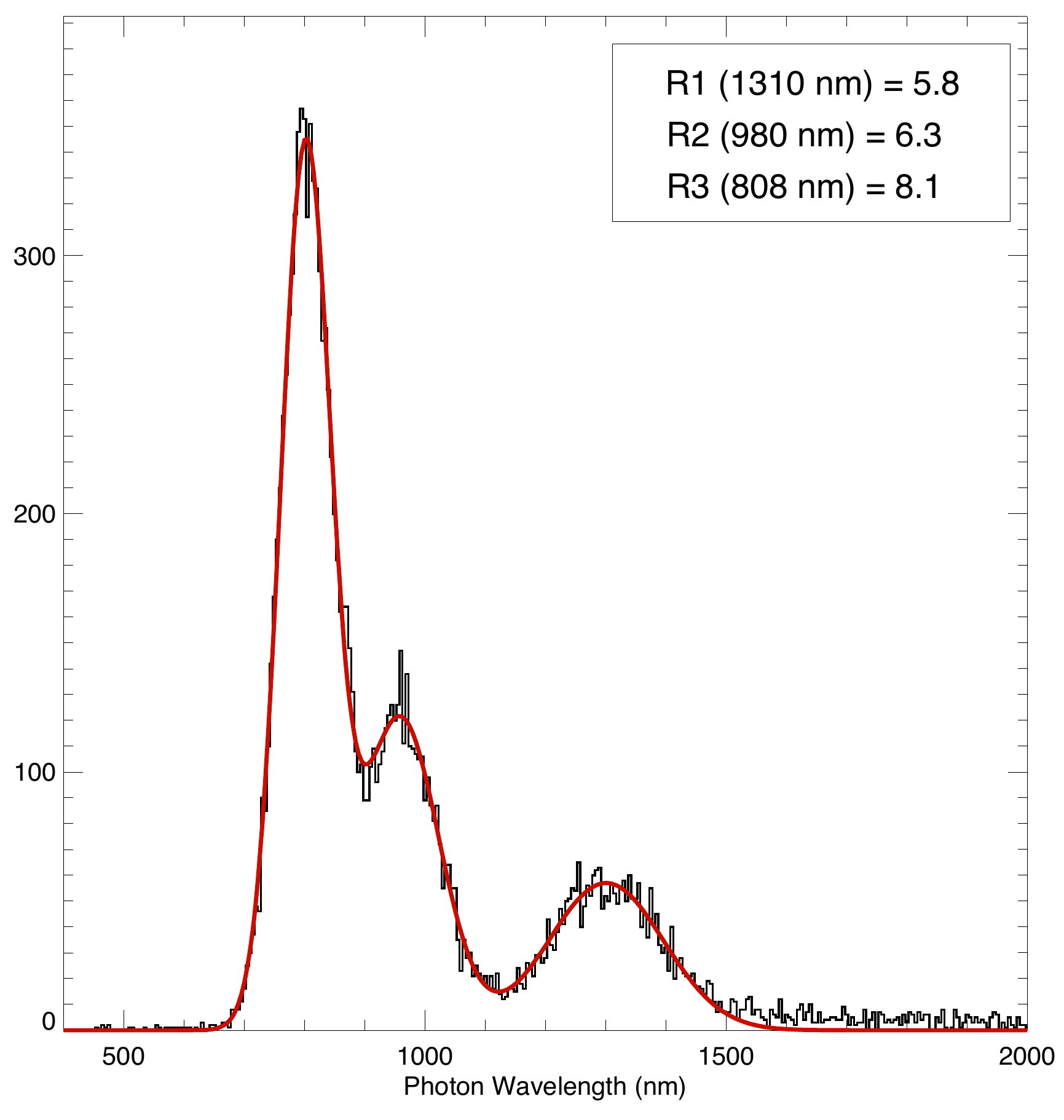}
\end{center}
\vspace{-.2in}
\caption{A histogram of pulse heights recovered using optimal filtering when a PtSi resonator of the design shown in Figure~\ref{fig:SCI4} is illuminated by 808, 920, and 1310 nm photons from laser diode sources.  A simultaneous gaussian fit of peaks gives a measured R of 8.1 at 808 nm.} 
\label{fig:measR}
\vspace{-.1in}
\end{figure}

The primary theoretical limitation on the spectral resolution is from the intrinsic quasiparticle creation statistics during the photon absorption event.  The energy from the photon can end up in two places, the quasiparticle system and the phonon system.  These systems interact, allowing energy exchange between the two, which reduces the statistical fluctuation from Poisson by the Fano factor $F$, typically assumed to be 0.2~\cite{Kozorezov:2008ii}. The spectral resolution of the detector, $R = \lambda/ \Delta \lambda =  E/\Delta E$, can be written as $R = \frac{1}{2.355}\sqrt{\frac{\eta h \nu}{F \Delta}}$, where $\eta$ is the efficiency of conversion of energy into quasiparticles, typically 0.57~\cite{Kozorezov:2007ek}, $h \nu$ is the photon energy, $F$ is the Fano factor, and $\Delta$ is the energy gap.  The energy gap depends on the superconducting transition temperature $(T_c)$ of the inductor, $\Delta \approx 1.72 k_B T_c$, and we typically operate at a base temperature of $T_{op}=T_c/8$.  Going to lower $T_c$, and hence lower operating temperature, improves the theoretical $R$.  Operating at 100 mK yields a theoretical spectral resolution of R=115 at 400 nm. The Fano Limit as a function of operating temperature (assuming the MKID is operated at T$_c$/8) is shown in Figure~\ref{fig:fano}.  Previous research with Superconducting Tunnel Junctions (STJs) with superconducting absorbers has shown that superconducting absorbers can approach the Fano limit~\cite{Li:2001dc,Huber:2004ga,leGrand:1998vh}.

After over a decade of development we are currently producing high quality PtSi MKID arrays.  These are the best UVOIR MKID arrays ever produced with 20,440 pixels, $>$90\% of the pixels functional, R $\approx$ E/$\Delta$E$\sim$8 at 808 nm~(Figure~\ref{fig:measR}), and a quantum efficiency of ${\sim}$35\%.  In the last several months, we have had success developing anti-reflection coatings that appear to boost the quantum efficiency (QE) of our PtSi films above 70\% (see \S\ref{sec:AR}), and future optimization should further increase this gain.  These state-of-the-art MKID arrays are discussed in detail in a recent paper in Optics Express~\cite{Szypryt:2017cb}.  

\subsection{Linear MKID Arrays}
\label{sec:linear}

\begin{figure}
\begin{center}
\vspace{-0.0in}
\includegraphics[width=0.95\columnwidth]{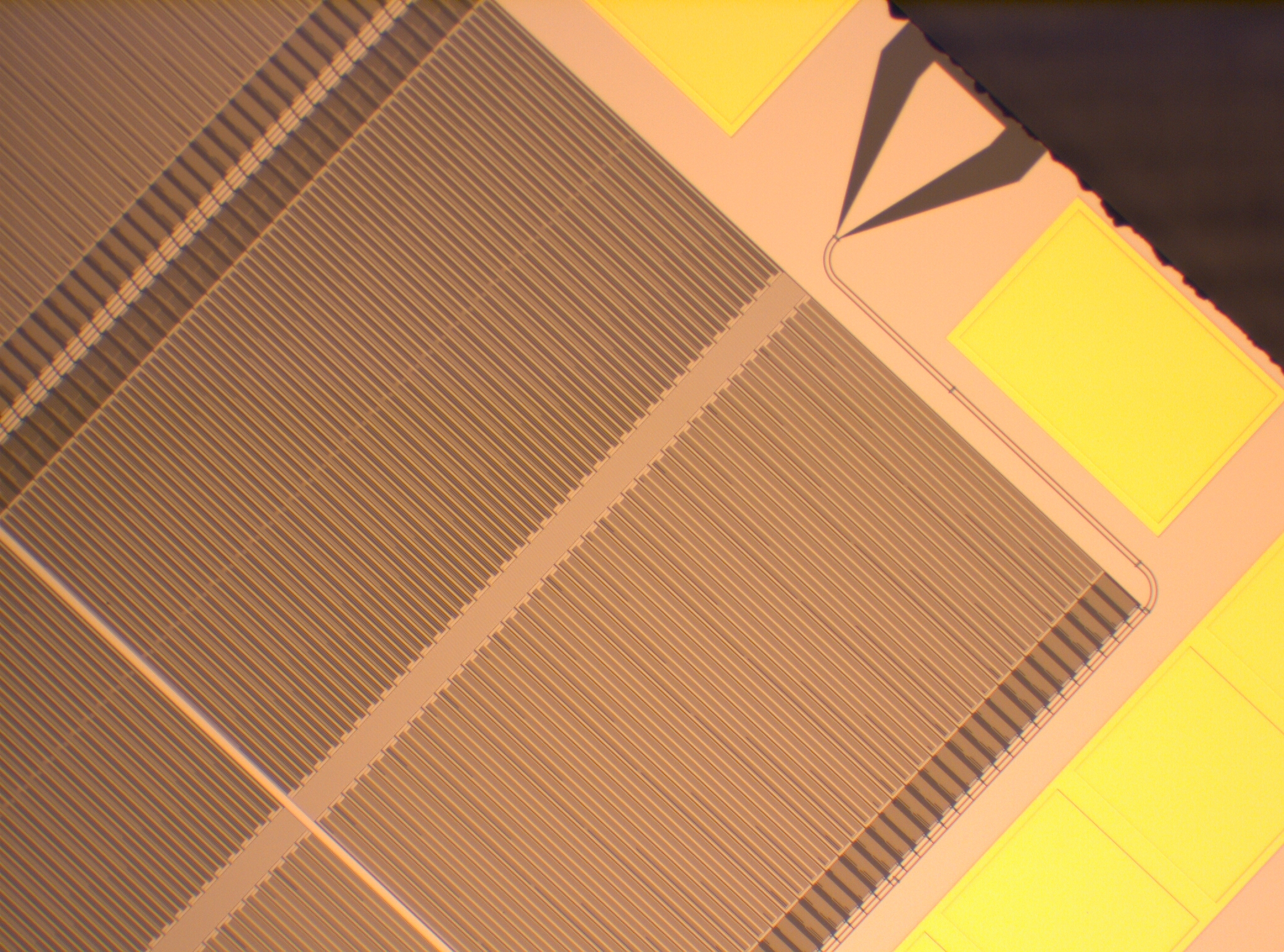}
\end{center}
\vspace{-.2in}
\caption{An optical microscope image of a linear PtSi MKID array.}
\label{fig:S1z}
\vspace{-0.1in}
\end{figure}

The most recent MKID breakthrough is the development of linear MKID arrays designed for spectroscopy, shown mounted in a microwave box in Figure~\ref{fig:S1}.  These arrays use the flexibility of the lumped element MKID geometry to put the photosensitive inductors in a linear strip, with the capacitor, coupler, and feedlines off to the side.  There is a 180 degree rotated version on the other side of the central strip.  This enables a photosensitive strip with a pixel pitch of 20~\micron~in the dispersion direction, and 60~\micron~in the cross dispersion direction, leaving 40~\micron~in the dispersion direction for each capacitor and coupler.  The strips are separated by 2 mm in the cross dispersion direction.  This design also obviates the need for a microlens array to boost the fill factor, and contains a larger capacitor volume which will lower TLS noise and improve energy resolution.  Figure~\ref{fig:S1z} shows a fabricated linear MKID array.

The first PtSi linear arrays have just been tested and have shown high internal quality factor (Q$_i$) resonances, although manufacturing defects in the feedlines have precluded a detailed analysis of their performance.  As both the design and manufacturing process is nearly identical to our MEC arrays we expect to have fully functioning, optimized, and AR coated arrays ready soon.  

\subsection{Anti-Reflection Coatings}
\label{sec:AR}

\begin{figure}
\begin{center}
\vspace{-0.0in}
\includegraphics[width=0.95\columnwidth]{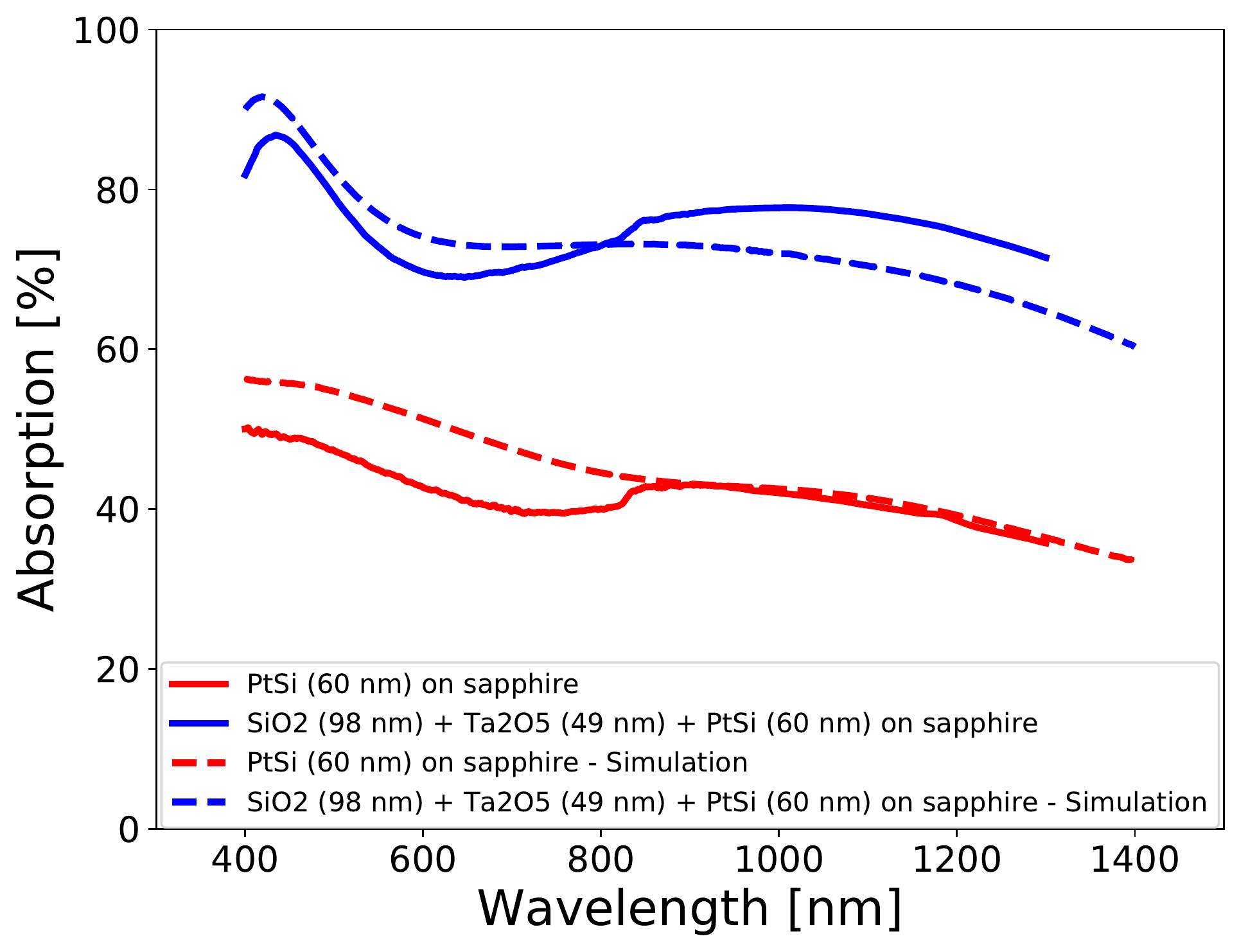}
\end{center}
\vspace{-.2in}
\caption{Absorption of a PtSi film deposited on a sapphire substrate in the $\SIrange{400}{1400}{\nm}$ wavelength range (Red) and absorption of the same film coated with the SiO\textsubscript{2}/Ta\textsubscript{2}O\textsubscript{5} bi-layer (Blue) - Dashed: simulations, Solid: measurements.}
\label{fig:plot_absportion}
\vspace{-.1in}
\end{figure}

The main goal of an anti-reflective (AR) coating is to lower the reflectance of an optical surface by creating destructive interference for the reflected light and constructive interference for the transmitted light. The performances of optical coatings are thickness sensitive and simulations can be carried out to tune the thicknesses of the different layers to optimize the absorption into the detector. Common materials for anti-reflection coating are SiO\textsubscript{2} ($n = 1.45$) and Ta\textsubscript{2}O\textsubscript{5} ($n = 2.16$) because their optical parameters are nearly constant over a large wavelength range, which allows nearly perfect impedance matching. The absorption of the PtSi ($\SI{60}{\nm}$)/SiO\textsubscript{2}/Ta\textsubscript{2}O\textsubscript{5} multi-layer is optimized with a software (TFcalc) in the $\SI{400}{}$ to $\SI{1400}{\nm}$ band by tuning the thicknesses of the two oxides. The best results, with thicknesses of $\SI{98}{\nm}$ and $\SI{49}{\nm}$ for the SiO\textsubscript{2} and Ta\textsubscript{2}O\textsubscript{5} respectively, are shown in Fig \ref{fig:plot_absportion}, showing an improvement in QE of nearly a factor of 2 to roughly 75\% across the $\SIrange{400}{1400}{\nm}$ wavelength range. 


Experiments in our lab have shown that  AR coatings, when applied only to the photosensitive inductor (using a liftoff stencil), do not affect inductor quality factor or TLS noise.  

\section{MKIDs in the 2020s}

In 2010 OIR MKIDs were functioning in the lab only as single pixel, single layer test devices illuminated by fibers with readouts capable of probing only two MKIDs at once.  In 2019, high quality 20 kpix MKID instruments with high efficiency optical coupling to the telescope are permanently attached to the Subaru Telescope doing regular observations of exoplanet and disks.  This excellent progress has been due to steady funding from NASA for the detectors and NSF for the instruments.  If this support continues and expands we expect the 2020s to be a critical decade in which MKIDs begin to outcompete conventional semiconductor detectors for design wins in general purpose astronomical instruments. This decade will also likely see the first MKID instruments selected for a NASA mission.  Several critical projects are needed to pave the way for this photon counting, energy resolving future.

\subsection{Improving MKID Fabrication}

MKID arrays are already doing useful science, but suffer from issues due to the research nature of the UCSB and JPL fabs that make them that could preclude their use in some applications, such as precision photometry from space.  These problems manifest as dead pixels from lithographic problems and particulate contamination, and even dead sectors of the array from microwave feedline defects.  The best arrays from UCSB and JPL currently have about 80\% of their pixels pass the most stringent tests, which is limiting for some science cases like precision time series photometry.  While this yield may seem low compared to commercial detectors in the UVOIR, it is similar to the yield achieved by other superconducting arrays made in research fabs, for example for sub-mm and CMB astronomy (for example, SCUBA-2 in Table 2 of Holland \etal~\cite{2013MNRAS.430.2513H}).   

One attractive option to develop science grade MKIDs is through a collaboration between UCSB and MIT Lincoln Laboratory (MIT/LL). UCSB is a leader in MKID development and characterization.  MIT/LL is a leader in the field of superconducting electronics from decades of investment in infrastructure and process development, with the best superconducting production fabrication facility in the world. Low temperature detector arrays like the microcalorimeters for X-ray astronomy have benefited from leveraging MIT/LL's superconducting electronics fabrication process for developing complex arrays~\cite{2019ITAS...2902530D}. It should be possible to increase the performance and yield of MKIDs by fabricating UCSB-designed MKIDs at MIT/LL production fabrication facility.  In fact, MIT/LL has all the equipment required to make UCSB-style OIR MKIDs already in their production line. 

\subsection{Improving MKID Spectral Resolution}
\label{sec:improve}
Most OIR MKID development in recent years has been targeted for AO-fed instruments working in J band, centered around 1.3~\micron.  The best spectral resolution achieved was roughly R${\sim}$8 at 1~\micron.  MKIDs work better at shorter wavelengths, and this demonstrated resolution should translate to R=15--20 at 400 nm.  While this resolution is sufficient for many science goals, applications like sorting orders from an echelle and achieving high resolution (\S\ref{sec:highspec}) is quite challenging and a higher spectral resolution is very desirable.  The Mazin Lab at UCSB has recently been funded by NASA for a 3 year program devoted to improving the spectral resolution of MKIDs using two very promising techniques.
 
The first technique is using a different superconductor and operating the MKIDs at a lower temperature. This increases the detector sensitivity and, for a fixed phase response, lets us make a higher volume detector that we can readout at higher power.  The quasiparticle lifetime also tends to increase at lower temperatures, improving R as the $\sqrt{\tau_{qp}}$.  We have recently seen R=10 at 808 nm using Hf MKIDs running at 20 mK.

The second technique is using a newly developed amplifier, the quantum-limited travelling wave parametric amplifier developed at JPL. We currently use high electron mobility (HEMT) amplifiers from Sander Weinreb at Caltech and the Low Noise Factory.  These semiconductor amplifiers have excellent performance, with noise temperatures between 2--4 K in the 4--8 GHz band.  This is, however, far from the standard quantum limit of $\hbar \omega /2$ of added noise power per unit bandwidth, which works out to a noise temperature of approximately 50 mK/GHz.  Recently developments in quantum limited parametric amplifiers are beginning to open up new possibilities.  Two main types of parametric amplifiers are being studied, based on Josephson junctions~\cite{2015Sci...350..307M} or non-linear inductance in transmission lines~\cite{HoEom:2012kq}.  The maximum readout power and dynamic range of the Josephson-based para-amps is far too low for an MKID array, leaving the transmission line amplifiers as the primary contender. We have recently tested one of our standard MEC arrays at JPL with this amplifier, and boosted R from 7 to 10 at 808 nm. Noise measurements show that we should have gotten R${\sim}$25 with this configuration (APL, submitted), so there is likely some pulse height dependance on photon absorption location inside our pixel that we are in the process of eliminating.

\begin{figure}
\begin{center}
\vspace{-0.0in}
\includegraphics[width=1.0\columnwidth]{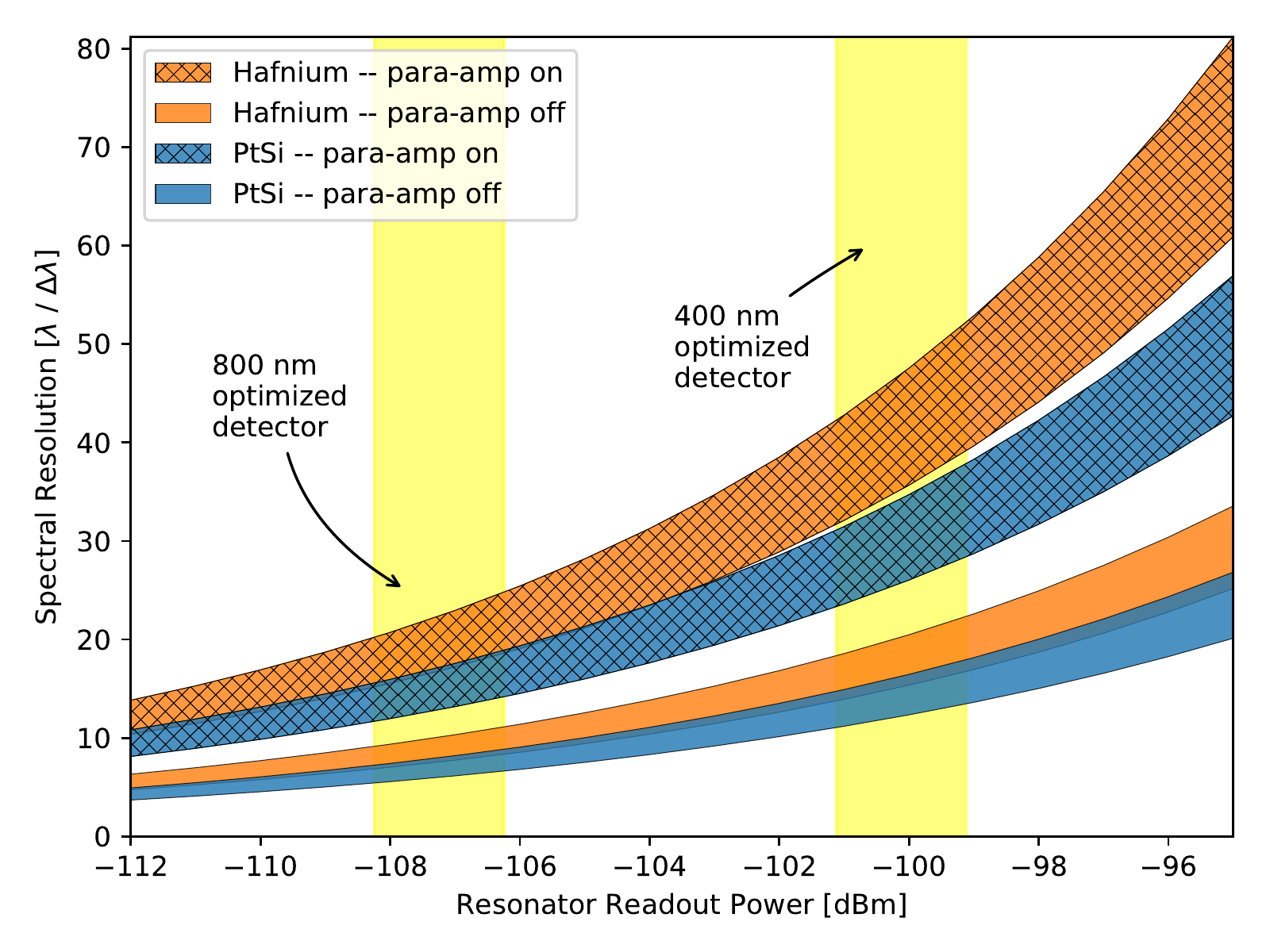}
\end{center}
\vspace{-.25in}
\caption{Plotted are projected detector spectral resolutions as a function of microwave power delivered to the hafnium and PtSi detectors with and without a parametric amplifier. Detectors optimized for different minimum wavelengths have different readout powers, and two regions of readout power are highlighted from past instruments (400 nm for the ARCONS array and 800 nm for the MEC array). For each material, measurements of the device noise are fit to a model for two-level systems and amplifier noise. The two-level-system noise in our detectors scales as $1/P^{1/2}$ and the amplifier noise scales as $1/P$. The spectral resolution for each power is calculated with the standard optimal filter formalism by rescaling the measured noise to the new power and using the material dependent pulse template.  Since this calculation does not include expected readout power enhancement from using Hf and the reduction on TLS noise expected in the larger S1 pixels it should be understood as our best guess at a lower limit of R$_{400}$ of roughly 30--50. }
\label{fig:rmodel}
\vspace{-.1in}
\end{figure}

With Hf MKIDs, optimized TLS loss through better surface preparation (also a priority for the superconducting quantum computation community), and parametric amplifiers OIR MKIDs could have spectral resolutions approaching R${\sim}$100 in the optical by the end of the 2020s.

\subsection{Improving MKID Readouts}

UCSB is currently developing the third generation of digital MKID readouts using the newly released and incredibly powerful Xilinx RFSOC, a single system-on-a-chip that contains 8 4 GSample/sec 12-bit ADCs and 16-bit DACs, a powerful FPGA, and an ARM microprocessor.  This Gen3 readout, which should be complete by 2020, promises to be able to read out 8192 MKIDs on a single readout board using only ${\sim}$10 Watts.  As powerful ADCs and FPGAs like the RFSOC continue to improve, MKID readouts will become cheaper and more powerful.  We project a Gen3 readout with donated FPGAs (Xilinx has been very generous about this) to cost less than \$1 per pixel, where the cryogenic HEMT is beginning to dominate the readout cost.  By the end of the decade MKID readouts for Megapixel class arrays will be common.

\subsection{Demonstrating MKID Order Sorting}

As described in \S\ref{sec:highspec} MKID echelle order sorting spectrographs have tremendous scientific potential.  In order to realize this potential significant work needs to be done to further develop the technology.  This work includes developing optimized linear MKID arrays, maturing MKID spectrograph designs, demonstrating the ability to sort orders in a lab testbed, and understanding how to optimally extract spectral information from the time-tagged photon lists.

\subsection{Funding Climate Disadvantages Transformative Instruments}

The funding situation for astronomical instruments in the US is difficult.  Simple estimates show Europe is outspending the US by a factor of 10 on instruments for 8--10-m telescopes (if we ignore LSST).  This extremely tight funding environment leads to conservative choices that disadvantage potentially transformative but less mature technologies like MKIDs.

\subsection{Broadening Participation in MKID Research}

Much of this whitepaper is centered around the OIR MKID work happening at UCSB with some fab work at JPL's Microdevice Lab (MDL).  This is an unfortunately side effect of UCSB being the only site in the US doing this kind of work.  Europe has recognized this opportunity, with new and large optical MKID programs started at U. Dublin (Prof. Tom Ray), U. Durham (Prof. Kieran O'Brien), SRON (Dr. Pieter de Vissers), and Observatoire de Paris (Dr. Faouzi Boussaha).  Durham and Paris have both been funded to build new MKID-based OIR instruments. At least one more OIR MKID node in the US is essential for stability and to maintain US leadership.  


\bibliographystyle{unsrt}
\bibliography{/Users/benmazin/Data/Projects/bib/mazin3.bib}

\end{document}